\documentclass[runningheads]{llncs}
\usepackage{multirow}
\usepackage{ulem}

\usepackage{xcolor}
\definecolor{bg}{HTML}{282828}
\usepackage{graphicx}
\usepackage{amssymb}
\usepackage{amsmath}
\usepackage{float}
\usepackage{caption}
\usepackage{subcaption}
\usepackage[colorlinks=true,linkcolor=cyan, citecolor=blue, urlcolor=teal]{hyperref}

\usepackage{listings}
\usepackage{color}
\usepackage{ulem}
\usepackage{tikz}
\usepackage{soul}
\usepackage{xspace}

\newcommand{\nate}{\textsc{Nate}\xspace}

\newcommand{\sherrloc}{\textsc{SHErrLoc}\xspace}
\newcommand{\logistic}{\textsc{Logistic}\xspace}

\newcommand{\bert}{\textsc{Bert}\xspace}
\newcommand{\codebert}{\textsc{CodeBert}\xspace}
\newcommand{\ocamlbert}{\textsc{OCamlBert}\xspace}

\newcommand*\circled[1]{\tikz[baseline=(char.base)]{
            \node[shape=circle,draw,inner sep=1pt] (char) {#1};}}

\definecolor{codegreen}{rgb}{0,0.6,0}
\definecolor{codegray}{rgb}{0.5,0.5,0.5}
\definecolor{codepurple}{rgb}{0.58,0,0.82}
\definecolor{backcolour}{rgb}{0.95,0.95,0.92}

\lstdefinestyle{mystyle}{
     commentstyle=\color{codegreen},
     keywordstyle=\color{magenta},
     numberstyle=\tiny\color{codegray},
     stringstyle=\color{codepurple},
     basicstyle=\footnotesize\ttfamily,
     breakatwhitespace=false,
     breaklines=true,
     captionpos=b,
     keepspaces=true,
     literate={<<}{\color{magenta}}1 %
             {<<r}{\color{dRed}}1 %
             {<*}{\color{dGreen}}1 %
             {<dim}{\color{dimgrey}}1 %
             {>>}{\color{black}}1,
     numbers=left,
     numbersep=5pt,
     showspaces=false,
     showstringspaces=false,
     showtabs=false,
     tabsize=2
 }

\makeatletter
  \newcommand\reduline{\bgroup\markoverwith{\textcolor{red}{\rule[-0.5ex]{2pt}{0.4pt}}}\ULon}
  \UL@protected\def\blueuwave{\leavevmode \bgroup 
    \ifdim \ULdepth=\maxdimen \ULdepth 3.5\p@
    \else \advance\ULdepth2\p@ 
    \fi \markoverwith{\lower\ULdepth\hbox{\textcolor{red}{\sixly \char58}}}\ULon}
  \UL@protected\def\yellowdotuline{\leavevmode \bgroup 
    \UL@setULdepth
    \ifx\UL@on\UL@onin \advance\ULdepth2\p@\fi
    \markoverwith{\begingroup
       \lower\ULdepth\hbox{\kern.06em \textcolor{yellow}{.}\kern.04em}%
       \endgroup}%
    \ULon}
  \UL@protected\def\greendashuline{\leavevmode \bgroup 
    \UL@setULdepth
    \ifx\UL@on\UL@onin \advance\ULdepth2\p@\fi
    \markoverwith{\kern.13em
    \vtop{\color{green}\kern\ULdepth \hrule width .3em}%
    \kern.13em}\ULon}
\makeatother

\begin{document}
\title{Novice Type Error Diagnosis with Natural Language Models}

\author{Chuqin Geng\inst{1,2} \orcidID{0000-0002-3563-1596}\and 
Haolin Ye\inst{1} \orcidID{0000-0002-7402-617X}\and
Yixuan Li\inst{1} \orcidID{0000-0001-9349-5476}\and
Tianyu Han\inst{1} \orcidID{0000-0001-6582-165X}\and
Brigitte Pientka\inst{1} \orcidID{0000-0002-2549-4276}\and
Xujie Si\inst{1,2,3} \orcidID{0000-0002-3739-2269} 
}

\institute{McGill University
\and Mila - Quebec Institute
\and CIFAR AI Research Chair\\
\email{\{chuqin.geng,haolin.ye,yixuan.li,tianyu.han2\}@mail.mcgill.ca\\
\{bpientka,xsi\}.cs.mcgill.ca
}\\
}
%
%

%
%

\maketitle              

\begin{abstract}

Strong static type systems help programmers eliminate many errors without much burden of supplying type annotations. 
However, this flexibility makes it highly non-trivial to diagnose ill-typed programs, especially for novice programmers.  
Compared to classic constraint solving and optimization-based approaches, the data-driven approach has shown great promise in identifying the root causes of type errors with higher accuracy. 
Instead of relying on hand-engineered features, this work explores natural language models for type error localization, which can be trained in an end-to-end fashion without requiring any features. 
We demonstrate that, for novice type error diagnosis, the language model-based approach significantly outperforms the previous state-of-the-art data-driven approach. Specifically, our model could predict type errors correctly 62\% of the time, outperforming the state-of-the-art \nate's data-driven model by 11\%, in a more rigorous accuracy metric.
Furthermore, we also apply structural probes to explain the performance difference between different language models.


\end{abstract}
\section{Introduction}
\definecolor{lightblue}{rgb}{.0,.95,1.05}
\label{sec: introduction}
Diagnosing type errors has received much attention from both industry and academia due to its potential of reducing efforts in computer software development. 
Existing approaches such as standard compilers with type systems, report type errors through type checking and constraint analysis. Thus, they merely point to locations where the constraint inconsistencies can occur and such locations might be far away from the true error source. Moreover, type error localization would require programmers to understand the functionality of type systems and check which part of the code contradicts their intent. Languages such as C and Java force programmers to write annotations which make the code neat. It also makes it easier to find the roots of type errors. Strongly typed functional languages such as OCaml and Haskell, however, programmers need not bother with annotations, since type systems automatically synthesize the types. The absence of type annotation comes at a price: novices could easily get lost in debugging their programs and locations of constraint inconsistencies from error messages can be misleading. 



Joosten et al~\cite{joosten} suggests that beginners usually pay more attention to  underlined error locations rather than error messages themselves when fixing programs. Therefore it is necessary to ameliorate the localizing performance of these type systems. Let us consider an OCaml ill-type program in \ref{fig: sumList program}. Although the programmer intends to write a function that sums up all numbers from a list, they mistakenly put the empty list, \texttt{[]}, at line 3 as a base case. This should instead be \sethlcolor{lightblue}\textcolor{black}{\hl{0}} as shown in Fig~\ref{fig: Fixed sumList program}. 
The compiler identifies the type error in line 5 saying that the head of the list, \texttt{h}, has type list rather than integer as required by the integer addition operator. 

\begin{figure}
     \centering
     \begin{subfigure}[b]{0.9\textwidth}
         \centering
        \begin{minipage}{13.2cm}
        \begin{lstlisting} [escapechar=\%]
        let rec sumList xs =
          match xs with
          | [] -> %\sethlcolor{yellow} \textcolor{red}{\hl{[]}}% (* root cause *)
          | h :: [] -> h
          | h :: t -> %\blueuwave{h}% + sumList t (* misleading complaint *)
    %\sethlcolor{red}\textcolor{white}{\hl{this expression has type 'a list but was expected of type int}}%
        \end{lstlisting}
        \end{minipage}
         \caption{an ill-typed OCaml program that aims to sum all the elements from a list}
         \label{fig: sumList program}
     \end{subfigure}
     \hfill
     \begin{subfigure}[b]{0.9\textwidth}
        \definecolor{lightblue}{rgb}{.0,.95,1.05}
       \begin{minipage}{14.2cm}
\begin{lstlisting} [escapechar=\%] %  numbers=none
        let rec sumList xs =
          match xs with
          | [] -> %\sethlcolor{lightblue} \textcolor{black}{\hl{0}}% (* <= correct fix *)
          | h :: [] -> h
          | h :: t -> h + sumList t 
\end{lstlisting}
\end{minipage}
    \caption{the fixed version of the OCaml code above}
         \label{fig: Fixed sumList program}
     \end{subfigure}
     \caption{A simple example of OCaml type error and its relevant fix.}
     \hfill
\end{figure}

    

This illustrates that programmer's intent plays an important role in localizing type errors. To tackle this issue, \nate~\cite{nate} proposes to use data-driven models to diagnose type errors. In this way, programmers' intent can be learned and incorporated into machine learning models. \nate~'s best model could achieve over 90\% accuracy in diagnosing type errors. Although this is an exciting result, \nate~'s models are evaluated with a rather loose metric and heavily rely on a considerable amount of hand-designed feature engineering. In addition, these features are designed in an ad-hoc fashion which prevents them from being directly applied to other language compilers.

Our approach adopts transformer-based language models to avoid considerable feature engineering. As we treat programs as natural language texts, these models do not rely on any knowledge or features about the specific programming language, thus they can be easily applied on any language. This method may seem to ignore the syntactic structure of a given programming language. However, we use structural probes~\cite{probe} to demonstrate the structure is embedded implicitly in the deep learning models' vector geometry in Section~\ref{sec: evaluation}. We also propose a more rigorous metric, and show language models outperform not only standard OCaml compiler and constraints-based approaches but also the state-of-the-art \nate~'s models under the new metric.

Transformer-based models have achieved great success in a wide range of domains in computer science including natural language processing. BERT~\cite{bert} and GPT~\cite{gpt-2,gpt-3}, popular transformer variants, have shown incredible capability of understanding natural languages. Together with its pre-training and fine-tuning paradigm, these models can transfer knowledge learned from a large text corpus to many downstream tasks such as token classification and next sentence prediction. Empirical results suggest that the performance of these language models even exceeds the human level in several benchmarks. In this work, we show how to take advantage of these powerful language models to localize type errors. First, we process programs as if they were natural language text and decompose the processed programs at the term or subterm level into token sequences so that they can be fed to language models. This allows us to turn the type error diagnosis problem into a token classification problem.
 In this way, language models can learn how to localize type errors in an end-to-end fashion.


\textbf{Contributions} We propose a natural language model-based approach to the type error localization problem. Our main contributions are as follows:\begin{itemize}
\item[$\bullet$] Without any feature engineering or constraints analysis, we apply different language models including BERT, CodeBERT, and Bidirectional LSTM to type error localization. 
\item[$\bullet$] We study training methodology such as positive/negative transfer to improve our models' performance. Instead of using a loose evaluation metric as proposed in previous work, we define a more rigorous, yet realistic, accuracy metric of type error diagnosis.
\item[$\bullet$] Empirical results suggest that our best model can correctly predict expressions responsible for type error 62\% of the time, 24 points higher than SHErrLoc and 11 points higher than the state-of-the-art \nate tool.
\item[$\bullet$] We study the interpretability of our models using structural probes and identify the link between language models' performance with their ability of encoding structural information of programs such as AST.
\end{itemize}

We start by presenting the baseline, our model architecture and the structural probe in Section~\ref{sec: approach}. Section~\ref{sec: dataset} introduces the dataset and evaluation metric, while Section~\ref{sec: evaluation} presents the experiential results and our discussion. Then, Section~\ref{sec: related work} gives an overview of related work. Finally, Section~\ref{sec: conclusion} concludes the whole paper and proposes some directions for future work.



\section{Approach}
\label{sec: approach}
In this section, we introduce deep learning-based language models including RNN, BERT, and CodeBERT. We take advantage of the pre-training and fine-tuning paradigm of language models and show how to transform the type error diagnosis problem to a token classification problem, a common downstream task in fine-tuning. We also present the structural probe which allows us to find the embedded structural information of programs from models' vector geometry.
\subsection{Language models} 
Deep learning has achieved great success in modelling languages since the invention of recurrent neural networks (RNNs) \cite{rnn}. RNNs adopt “internal memory” to retain information of prior states to facilitate the computation of the current state. Unlike traditional deep neural networks, the output of RNNs depends on the prior elements within the sequence which make them ideal for processing sequential inputs such as natural languages and programs.

In this study, we also choose a bidirectional long-short term memory (Bidirectional LSTM) \cite{bilstm} as our baseline model. However, RNNs are known to have several drawbacks such as a lack of parallelization and weak long-range dependencies. These two limitations are later addressed by the self-attention mechanism introduced by the transformer. Self attention~\cite{attention} is an attention mechanism relating different positions of a single sequence in order to compute a representation of the sequence. Transformers also follow an encoder-decoder architecture as other successful neural sequential models. Both its encoder and decoder have been studied and shown great capabilities for modelling natural languages and solving many downstream tasks.

BERT, which stands for Bidirectional Encoder Representations from Transformers takes advantage of the encoder part of the transformer while the GPT-n series are based on the decoder. In this work, we focus on BERT rather than GPT-3 \cite{gpt-3} for several reasons. First, BERT requires a fine-tuning process which alters the pre-trained model for specific downstream tasks. This fits our formalization of treating type error diagnosis as a downstream task. Second, the size of GPT-3 is enormous compared to BERT, making it hard to train and infer. Third, BERT is an open-source tool and easily available for users to access while GPT-3 is not open-sourced. 

\subsection{The pre-training and fine-tuning scheme}
The pre-training and fine-tuning scheme allows machine learning models to apply knowledge gained from solving one task to different yet related tasks. Compared to fine-tuning, pre-training is more essential as it determines what knowledge is learned and stored in machine learning models. As a result, there are some recent works on improving the pre-training scheme of language models.

BERT stands out by proposing two critical unsupervised tasks during pre-training - Masked Language Modeling (MLM) and Next Sentence Prediction (NSP)~\cite{bert}. MLM requires the model to predict masked-out tokens conditioned on other tokens within sentences whereas NSP forces the model to predict if the input two sentences are next to each other in the original document. These two training tasks or objectives allow the model to understand the natural language from a statistical perspective, and empirical results of BERT have shown that pre-training on large text corpus using these two objectives facilitates a wide range of downstream tasks.

\subsection{Type error diagnosis as token classification} 

\begin{figure}[h]
\centerline{\includegraphics[scale=0.4]{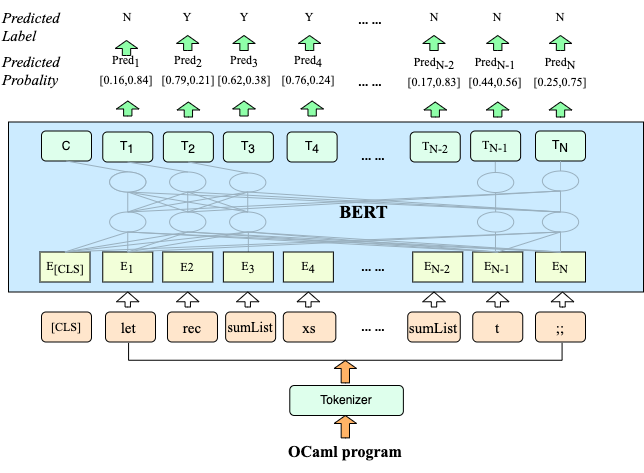}}
\caption{The type error diagnosis as a token classification task. After an input program is split into a sequence of tokens by a tokenizer, each token is fed into BERT to get an embedding representation. A simple classification head takes each token representation and outputs a predicted probability which indicates the model's belief of the current token being related to type error.}
\label{fig: bertcls}
\end{figure}

Token classification~\cite{bert} is a downstream task which uses a pre-trained Bert model with a token classification head on top to make a prediction for each token in a sentence. One of the most common token classification tasks is Named Entity Recognition (NER). The goal of NER is to find a label for each entity in a sentence, such as a person, location or organization.
Type error diagnosis can be naturally viewed as a token classification problem.
Note that type error diagnosis attempts to find type error locations within  a piece of code, so we can reformulate it to a token classification task if we assign label \textit{1} to all the tokens that contribute to the type error and label \textit{0} to those tokens that are unrelated to type error. Figure \ref{fig: bertcls} gives an overview of using token classification to achieve type error diagnosis.
As a fine-tuning task, token classification requires labelled data to provide ground truth to help the model learn. In the context of type error diagnosis, this means we need to have a dataset consisting of many ill-typed programs along with their true type error locations. We will discuss our dataset more in Section~\ref{sec: dataset}.  

Given that large language models are extremely expensive to train, even for industrial companies, a common practice is to fine-tune pre-trained large language models on a new dataset. In our case, we choose different configurations of BERT to explore the optimal model for type error diagnosis. Our models are as follows:\begin{itemize}
\item \textbf{Bidirectional LSTM}: The model is trained directly on the fine-tuned dataset from scratch. This model serves as our baseline.
\item \textbf{BERT from scratch}: To compare with Bidirectional LSTM, we train the BERT to do token classification from scratch, without any pre-training process. 
\item \textbf{BERT Small, BERT Medium, BERT Base, and BERT Large}: \newline As the name suggests, these four models are different in terms of size. Although they are pre-trained on the same dataset, we hypothesize that the size would affect the representation power of models and therefore would affect the performance of type error diagnosis. 
\item \textbf{CodeBERT}: CodeBert~\cite{codebert} is a pre-trained bimodal model for programming language (PL) and natural language (NL). It is pre-trained on several programming languages including Ruby, Javascript, Go, Python, Java, and PHP. As it is pre-trained on such programming languages that ask programmers to specify the type, we postulate that it may not work well on OCaml programs. However, we are still curious to see if these programming languages may share some patterns with OCaml which can enhance its error localization ability during the fine-tuning process.
\item \textbf{BERT pre-trained on OCaml}: Since BERT is pre-trained on natural language texts which do not contain OCaml programs,  we collect two datasets of OCaml programs, one from industry and the other one from students' homework submission. Then we pre-train BERT Base and BERT Large on them with the same training objectives. This technique is also called domain shift which could help the model perform better on downstream tasks which have different data distribution from that of the original input. Together with CodeBERT, these models allow us to explore how domain shift affects models' performance.
\end{itemize}

\begin{figure}[htbp]
\centering
\includegraphics[width=\linewidth]{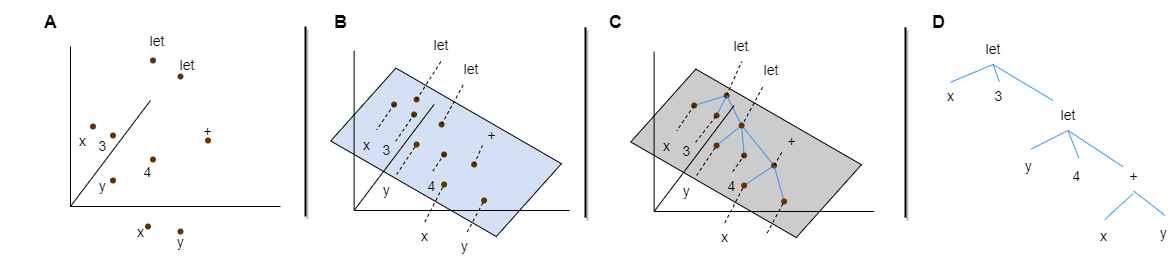}
\caption{Syntax tree of example OCaml program ``let x = 3 in let y = 4 in x + y''}
\label{fig: TREE}
\end{figure}

\subsection{Structural probe}
We attempt to use the structural probe method~\cite{probe} to find structural information embedded implicitly in the deep learning models' vector geometry.

In deep learning, each token has a vector representation after feeding into the model. The method finds a distance metric that can approximate the result of the distance metric defined by the syntax tree from applying to any two tokens of a program. More specifically, it defines a linear transform of the space in which squared $L_2$ distance between vectors best reconstructs tree path distance, and thus the structure of the tree is demonstrated by the geometry of the vector space. Figure \ref{fig: TREE} gives an overview of using the structural probe to reconstruct tree structure information of programs.

\section{Dataset and evaluation metric}
\label{sec: dataset}
In this section, we present the datasets and the evaluation metric that we use in our work. The training datasets that we use are the same ones used in the baseline. However, we propose a different metric of accuracy. Notice this metric has been applied to every model, so data produced by \nate's models may look different from their original paper. To explore the capability of language models, we also create two pre-training datasets consisting of over 370,000 OCaml programs in total.

\begin{table}
\centering
\begin{tabular}{|cc|r|r|c|c|} 
\hline
\multicolumn{2}{|l|}{}           & \multicolumn{1}{c|}{Num. of programs} & \multicolumn{1}{c|}{\begin{tabular}[c]{@{}c@{}}Average num. \\of tokens\end{tabular}} & \begin{tabular}[c]{@{}c@{}}Has ground-\\truth label\end{tabular} & Usage                                                                                          \\ 
\hline
\multirow{2}{*}{NATE} & SP14     & 2,712                                 & 136                                                                                   & Yes                                                              & \multirow{2}{*}{\begin{tabular}[c]{@{}c@{}}Fine-tuning \\(training and testing)\end{tabular}}  \\
                      & FA15     & 2,365                                 & 133                                                                                   & Yes                                                              &                                                                                                \\ 
\hline
\multicolumn{2}{|c|}{GitHub}     & 350,000                               & 121                                                                                   & N/A                                                              & \multirow{2}{*}{Pre-training}                                                                  \\
\multicolumn{2}{|c|}{Homework}   & 20,000                                & 99                                                                                    & N/A                                                              &                                                                                                \\
\hline
\end{tabular}
\vspace{5pt}
\caption{Statistics of pre-training and fine-tuning datasets}
\label{tb:BERTACC}
\end{table}

\subsection{Pre-training dataset}
The pre-training procedure plays an important role in transformer-based language models. The purpose of pre-training is to train the model on large-text corpus in an unsupervised fashion. After pre-training, models should have weights that encode the probabilities of a given sequence of words occurring in sentences. The success of modern language models is often attributed to large pre-training datasets. Motivated by this observation, we collect the following two datasets.
\begin{itemize}

\item \textbf{GitHub-OCaml dataset}  {Based on the default ranking configuration provided by GitHub, we collected top-500 OCaml GitHub repositories, which has been identified as an instance of ``fair use"~\cite{Keefe}. Given that most
samples of our GitHub dataset are from the industry, we perform some post-processing work by filtering out programs that are automatically generated by lexer and parser. Even though, some of them are still quite different from programs written by novice programmers. The resulting dataset contains over 350,000 OCaml programs.
}

\item \textbf{Student-OCaml dataset} {To collect OCaml programs written by novice programmers, we collected around 2, 000 homework submissions made by over 350 students from an undergraduate programming languages course taught at McGill University. Each homework submission consists of 10 subtasks (e.g., functions for a specific coding question) on average. This gives us 20,000 OCaml data samples.}
\end{itemize}

In Section~\ref{sec: evaluation}, we study how pre-training on these two datasets affects the performance of type error diagnosis.

\subsection{Fine-tuning dataset}
Fine-tuning, on the other hand, requires labelled data as supervision to facilitate model learning. This means we need a set of ill-typed programs along with the correct locations of type errors as ground truth. Manual ground truth annotation and ill-typed program collection can be troublesome.
Fortunately, \nate provides a dataset consisting of 5,000 labelled programs that cover many types and locations of the errors that beginners make in practice, together with the corresponding fixes.

\nate's dataset was collected from an undergraduate Programming Languages course at UC San Diego in Spring 2014 and Fall 2015, which are named SP14 and FA15 respectively. Besides providing supervision, \nate's dataset can also be used as a test-bed so we can compare our models with \nate's fairly.

\definecolor{lightblue}{rgb}{.0,.95,1.05}
\definecolor{lightgreen}{rgb}{0.79,2.50,0.79}
\definecolor{lightpurple}{rgb}{0.64,0.93,0.30}

\begin{figure}
     \centering
     \begin{subfigure}[b]{0.99\textwidth}
        \begin{lstlisting} [escapechar=\%]
let rec clone x n = if n <= 0 then [] 
                    else x :: (clone x (n - 1))
let rec helper x = if x = 0 then 1 else 10 * helper (x - 1)
let padZero l1 l2 =
  if (List.length l1) < (List.length l2)
  then ((%\sethlcolor{yellow} \textcolor{red}{\hl{clone}} % "0" %\sethlcolor{yellow} \textcolor{red}{\hl{List.length}}% l2) - (List.length l1)) :: l1
  else ((%\sethlcolor{yellow} \textcolor{red}{\hl{clone}}% "0" List.length l1) - (List.length l2)) :: l2
\end{lstlisting}
    \vspace{-10pt}
         \caption{An ill-typed OCaml program. Highlighted tokens on line 6 and 7 are predictions made by \nate's model.}
         \label{fig: Buggy program}
     \end{subfigure}
     \hfill
     \begin{subfigure}[b]{0.99\textwidth}
        \definecolor{lightblue}{rgb}{.0,.95,1.05}
        \begin{lstlisting} [escapechar=\%]
let rec clone x n = if n <= 0 then []
                    else x :: (clone x (n - 1))
(*let rec helper x = if x = 0 then 1 else 10 * helper (x - 1) *)
let padZero l1 l2 =
  if (List.length l1) < (List.length l2)
  then %\sethlcolor{lightblue} \textcolor{black}{\hl{(((clone 0 ((List.length l2) - (List.length l1))) @ l1), l2)}}%
  else %\sethlcolor{lightblue} \textcolor{black}{\hl{(((clone 0 ((List.length l1) - (List.length l2))) @ l2), l1)}}%
\end{lstlisting}
\vspace{-10pt}
    \caption{Fixed version of the OCaml program above. Highlighted tokens on line 6 and 7 form the ground truth, whereas the change on line 3 does not.}
    \label{fig: Fixed program}
     \end{subfigure}
     \caption{An ill-typed OCaml program and its corresponding fix.}
     \hfill
    \label{fig: Example program}
\end{figure}

\subsection{Evaluation metric}
\label{sec: eval metric}
\nate processes programs to sub-tree sets and then filters the sub-trees that are not related to type errors to get true error locations --- the ground truths. As a consequence, if a large sub-tree, T, is the ground truth, many of its sub-trees are treated as true error locations as well. Then, if a model predicts any of these sub-trees, the prediction accuracy would be 100\% under the \nate's metric. We illustrate this using an example as shown in Fig~\ref{fig: Example program}. As we can see in~\ref{fig: Buggy program}, the model blames the token, \sethlcolor{yellow}\textcolor{red}{\hl{clone}}, which happens to match the error token, \sethlcolor{lightblue} \textcolor{black}{\hl{clone}}, in~\ref{fig: Fixed program}. It is not easy to see why \sethlcolor{lightblue} \textcolor{black}{\hl{clone}} is an error location because we only highlighted the union of all error spans. It ends up to be one since it is a strongly-related subtree of the ground truth highlighted in blue at line 6. Therefore, the prediction matches to the ground truth, resulting in an accuracy of 100\% under \nate's metric. However, such prediction is merely a tiny portion of the union of all error spans highlighted in blue which makes this an over-evaluation.

As a result, \nate's metric overestimates the prediction accuracy of not only its own machine learning models but also our language models. To visualize the over-evaluation from a data point of view, we test \nate's models and some of our models under \nate's metric.
We use BERT Small, Base and Large models, and they achieve Top-3 accuracies of 80\%, 84\% and 87\% respectively. Generally speaking, they are comparable to \nate's models, whose accuracies range from 84\% to 90\%.

We solve the overestimation issue by treating programs as consecutive token sequences rather than trees. Hence by counting the number of correctly predicted tokens, we can get a more precise and strict accuracy between 0\% and 100\% rather than just 0 (miss) or 1 (hit). To be more specific, our models estimate the probabilities of type error blame for each token in a binary classification setting. By converting predicted probabilities to label 0 or 1 using a default threshold value of 0.5, gives us a collection of predicted token sequences, $P$. By transforming the ground truth denoted by $L$ to token sequences as well, the correctly predicted token set is simply the intersection of them, $P \land L$. 
However, a trivial prediction which simply predicates each token as type error, i.e. $P$ = $\{1,1,1,...,1,1\}$, could achieve 100\% accuracy due to $P \land L = L$. To prevent this from happening, we divide the size of $P \land L$ by that of $P \lor L$.

$$ Accuracy(P,L) = \frac{|P \land L|}{|P \lor L|}$$

\section{Evaluation}
\label{sec: evaluation}

In this section, we empirically evaluate several approaches to type error diagnosis with particular focus on the following research questions:\footnote[1]{Our artifact is available at~\cite{geng_chuqin_2022_7055133}.}

\noindent
\textbf{RQ1}: How well do language models and other baseline methods perform on type error diagnosis? \\
\textbf{RQ2}: To what extent do model size and transfer learning affect models' performance? \\
\textbf{RQ3}: How well do language models generalize to unseen data?\\
\textbf{RQ4}: Does the model's ability of encoding structure information contribute to prediction accuracy? \\


\noindent
\textbf{Implementation and training}. We implement our experiments using PyTorch, Tensorflow and HuggingFace library. We use a batch size of 32 for both pre-training and fine-tuning processes. For pre-training, we pre-train BERT Base and BERT Large on both pre-training datasets for 10 epochs. For fine-tuning, all BERT models are fine-tuned on \nate's training dataset for 30 epochs. We set the initial learning rate to 0.00003 and use a scheduler to alter the learning rate during fine-tuning. To avoid stochasticity, we run each experiment three times and take the average. All our models are trained on a Tesla P100-PCIE-16GB GPU.

\noindent
\textbf{Configurations of language models}.
We explore different configurations of BERT to find the best model. We call BERT Base and BERT Large (BERT+) pre-trained on the homework OCaml dataset OCamlBERT Base (OBERT) and OCamlBERT Large (OBERT+). As for the BERT Base pretrained on the GitHub OCaml dataset, we call it BERT pre-GitHub (PBERT). We also trained a BERT Base from scratch without leveraging pre-trained weights and name it BERT Init (IBERT).

\noindent
\textbf{Baselines}. We compare our language model-based approach with three baselines as follows:
\begin{itemize}
\item[$\bullet$] \textsc{OCaml}, which extracts the type error location from the error message from the standard OCaml compiler. It is worth noting that the standard OCaml compiler stops compiling immediately when any type check fails and thus cannot report multiple errors.
\item[$\bullet$] \textsc{SHErrLoc}, which identifies the minimum set of locations to patch a type error using Bayesian inference~\cite{sherrloc}.
\item[$\bullet$] \textsc{Nate}, which predicts the top-K most likely ASTs that contribute to the type errors based on 282 hand-designed features~\cite{nate}. 
Specifically, \textsc{Nate} uses five different machine learning models --- logistic regression (\textsc{Logistic}), decision tree (\textsc{Tree}), random forests (\textsc{Forest}), and two multi-layer perception models (\textsc{MLP-10} and \textsc{MLP-500}) with a single hidden layer of 10 and 500 neurons, respectively.  

\end{itemize}



\subsection{Performance of different models (RQ1)}

\begin{figure}[htbp]
\centerline{\includegraphics[width = 0.92\linewidth,scale=0.5]{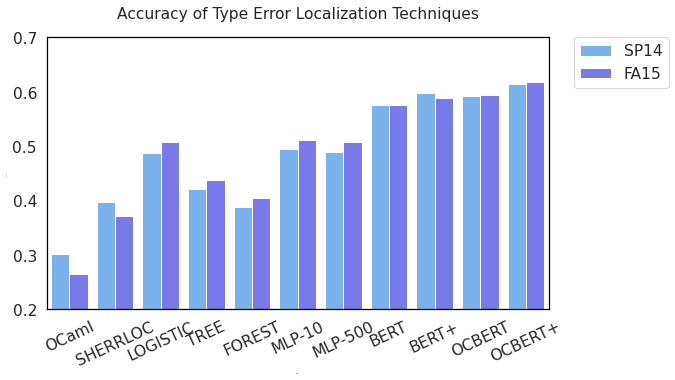}}
\caption{Comparison of accuracy of type error diagnosis methods.}
\label{fig:overall_accuracy}
\end{figure}

Fig~\ref{fig:overall_accuracy} summarizes our main results.  
We observe that both optimization-based approaches like \sherrloc and data-driven based approaches like \nate and various language models (i.e., BERT, BERT+, OCBERT, OCBERT+) outperform the standard OCaml compiler in terms of localizing root causes of type errors. 
Furthermore, data-driven approaches generally outperform optimization-based approaches, indicating that data plays a more important role compared to the pure optimization algorithm as adopted by \sherrloc. 
Among the five models used by \nate, it is a bit surprising that \logistic achieves similar performance as multi-layer perceptrons. We believe this is due to the rich hand-designed features which make simple models like logistic regression very effective. 

Nevertheless, we observe that our language model-based approaches significantly outperform \nate, which suggests that the embeddings learned in an end-to-end fashion are more effective than hand-designed features.

\begin{figure}[htbp]
\centerline{\includegraphics[width=\linewidth, scale=0.55]{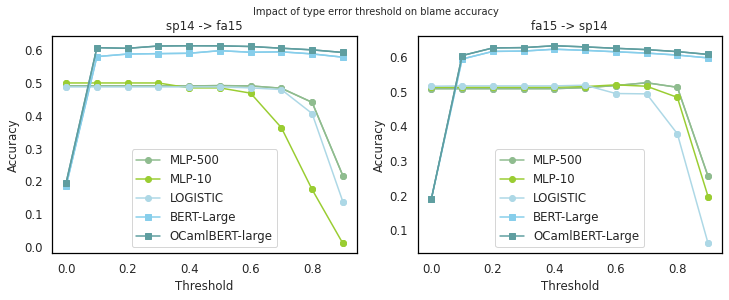}}
\caption{Impact of different thresholds on accuracy. }
\label{fig: threshold}
\end{figure}

Both \nate's and BERTs' output can be interpreted as a probability. Normally, we set the threshold to be 0.5, so if the output probability is greater than 0.5, the prediction will be 1, and 0 otherwise. The change in accuracy of models along with varying thresholds is reported in Fig~\ref{fig: threshold}. We notice that if we increase the threshold, BERTs' accuracy is robust whereas \nate drops significantly. This suggests that BERT models are much more confident with their predictions compared to \nate.

\subsection{Effectiveness of model sizes and transfer learning (RQ2)}

\begin{figure}[htbp]
\centerline{\includegraphics[width=\linewidth, scale=0.55]{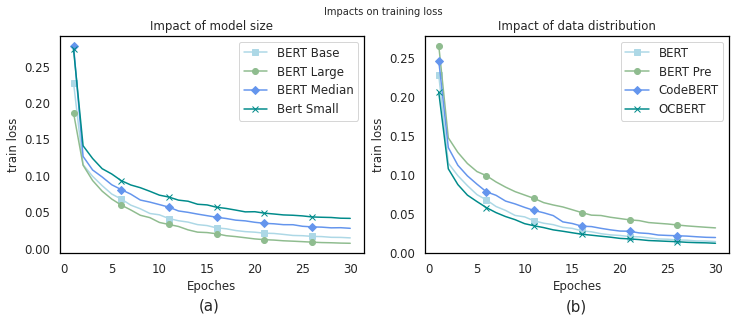}}
\caption{Impact on training loss. (a) shows how model size affects training loss while (b) illustrates how data distribution affects training loss. }
\label{fig: training loss}
\end{figure}

\textbf{Larger model leads to higher accuracy (not overfitting)}.
To study the effectiveness of different model sizes, we evaluate four modes of different sizes  --- Small (L=4, H=256), Medium (L=8, H=512), Base (L=12, H=768), and Large (L=16, H=1024). 
Fig~\ref{fig: training loss}(a) presents training loss curves of the four models of different sizes. This is somewhat expected since larger models usually tend to have lower training loss but may have an overfitting concern. This then may not lead to better accuracy. 

We further evaluate the testing accuracies of models of different sizes, which is summarized in Table~\ref{tb:BERTACC}.
The top half of Table~\ref{tb:BERTACC} shows the testing accuracies of four \bert models on four different train/test setups. The accuracy increases consistently when the model size increases on all train/test setups. This is very interesting because the train/test dataset is fixed with only the model size increasing, that is, with the same dataset, the larger model usually leads to higher accuracy instead of overfitting.



 \textbf{Positive/Negative transfer of learning} 
To study this objective, we focus on BERT, BERT pre-GitHub, CodeBERT, and OCamlBERT. Fig~\ref{fig: training loss}(b) presents the training loss curves of these four models. Since OCamlBERT is pretrained on only 20k code samples whereas BERT pre-GitHub is pretrained on 350k samples, it is quite surprising to notice that OCamlBERT has the lowest training loss whereas BERT pre-GitHub has the highest one. This is kind of counter-intuitive as models usually perform better when trained on a larger dataset. 
The bottom part of Table~\ref{tb:BERTACC} shows the testing accuracies of four \bert models on four different train/test setups. BERT pre-GitHub reports 51\% accuracy, 8 points lower than OCamlBERT which is consistent with training loss curves. The difference in accuracy could be explained by the positive/negative transfer of learning effect. As OCamlBERT is pretrained on code samples written by students/novice programmers (although from different universities working on different programming assignments), the similar data distribution in the fine-tuning process affects positively on accuracy \cite{transfer}. On the other hand, transfer learning impacts CodeBERT's and BERT pre-GitHub's accuracy due to disparate data distribution.

\begin{table*}[htbp]
\centering
  \begin{tabular}{|l|rrrr|}
    \hline
     & \ SP14/SP14 
     & \ FA15/FA15 
     & \ SP14/FA15 
     & \ FA15/SP14 \\
     & \ (Acc)
     & \ (Acc)
     & \ (Acc)
     & \ (Acc)\\
    \hline
    \bert Small & 68.83 & 63.05 & 50.45  &51.37 \\
    \bert Medium  & 71.53 & 65.08 & 53.39  &53.52 \\
    \bert Base  & 74.10 & 69.89 & 57.62 & 57.52  \\
    \bert Large  & 77.57 & 74.36 & 59.72 & 58.89  \\
    \hline
    Bi-LSTM  & 44.15 & 40.51 & 7.25  & 8.79 \\
    \bert Init  & 60.70 & 57.02 & 45.37  &43.98 \\
    \bert pre-GitHub & 68.52 & 59.59 & 51.84 & 51.43  \\
    \codebert  & 71.94 & 69.35 & 56.40  & 55.98 \\
    \ocamlbert Base  & 74.72 & 70.11 & 59.24 & 59.34  \\
    \ocamlbert Large  & 78.76 & 74.78 & 61.40 & 61.84  \\
    \hline
  \end{tabular}
  \vspace{5pt}
  \caption{Accuracies of different models evaluated on four train/test setups.}
  \label{tb:BERTACC}
\end{table*}




\subsection{Generalization ability of language models (RQ3)}
The generalization ability is an important property of our models as it measures how well a trained model performs on unseen program questions \cite{generalization}. 

To study this property, we focus on accuracy drops when evaluating different program problem sets, for instance, training on SP14 yet evaluating on FA15. We calculate accuracy drops using the difference between the first and last two columns of Table~\ref{tb:BERTACC}. We observe that relatively simple language models such as the Bidirectional LSTM model (Bi-LSTM) experience a large accuracy drop of over 30\% on unseen data. In addition, it achieves only 7.25\% and 8.79\% accuracies on the generalization tasks, which makes it almost useless in the real world. In contrast, the severest accuracy drop of BERTs accuracy is merely 17\%. This indicates that BERTs may have learned more robust and critical features which facilitate localizing type errors compared to Bi-LSTM.

In short, we should always consider large and powerful language models rather than small and simple ones when solving difficult tasks such as type error diagnosis.



\begin{figure}[tbp]
    \centering
    \begin{subfigure}[b]{0.45\textwidth}
         \centering
         \includegraphics[width =\textwidth]{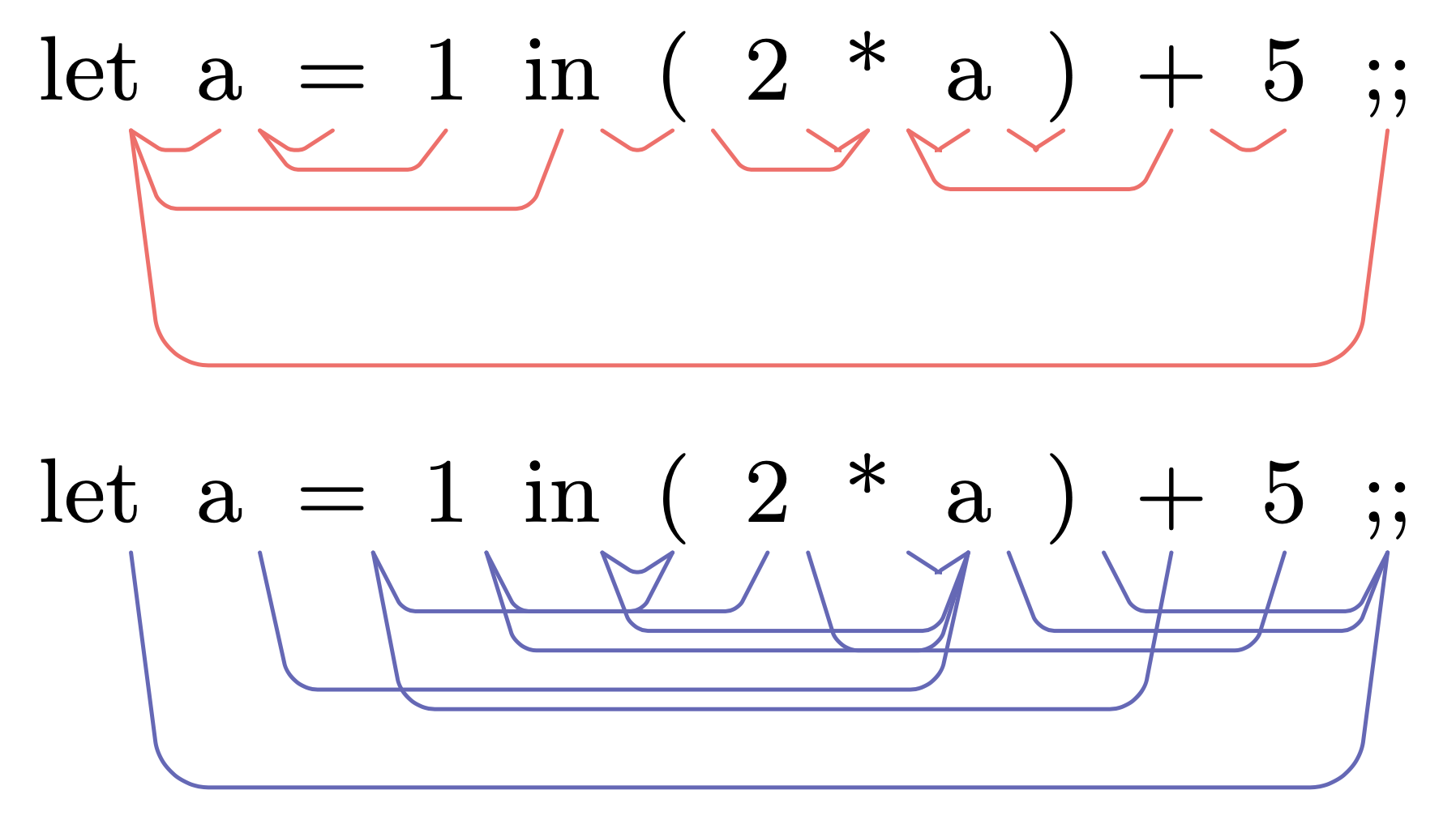}
         \caption{}
         \label{fig: ast_1}
    \end{subfigure}
    \hfill
    \begin{subfigure}[b]{0.52\textwidth}
         \centering
         \includegraphics[width = \textwidth]{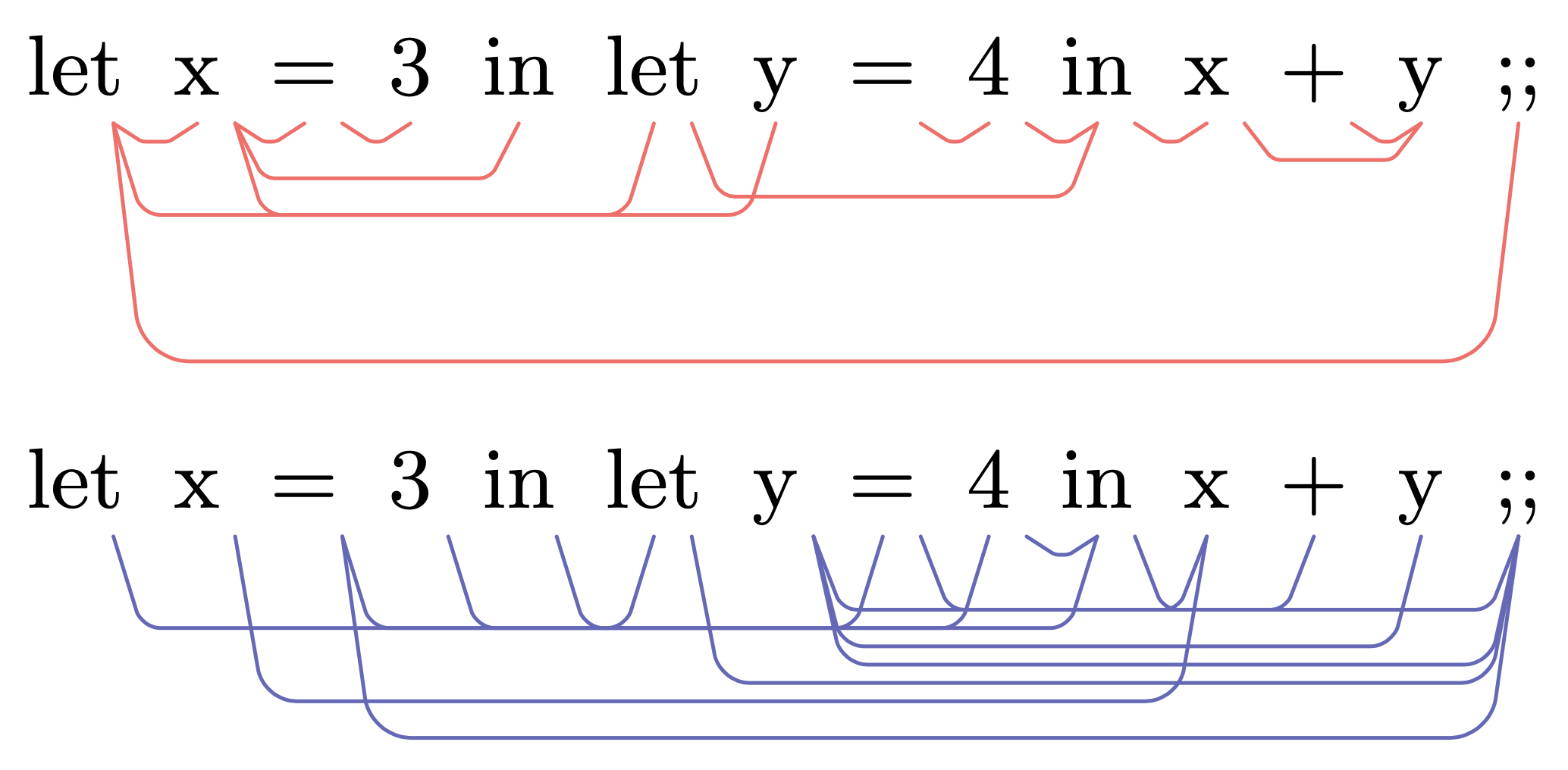}
         \caption{}
         \label{fig: ast_2}
    \end{subfigure}
    \begin{subfigure}[b]{0.7\textwidth}
         \centering
         \includegraphics[width = \textwidth]{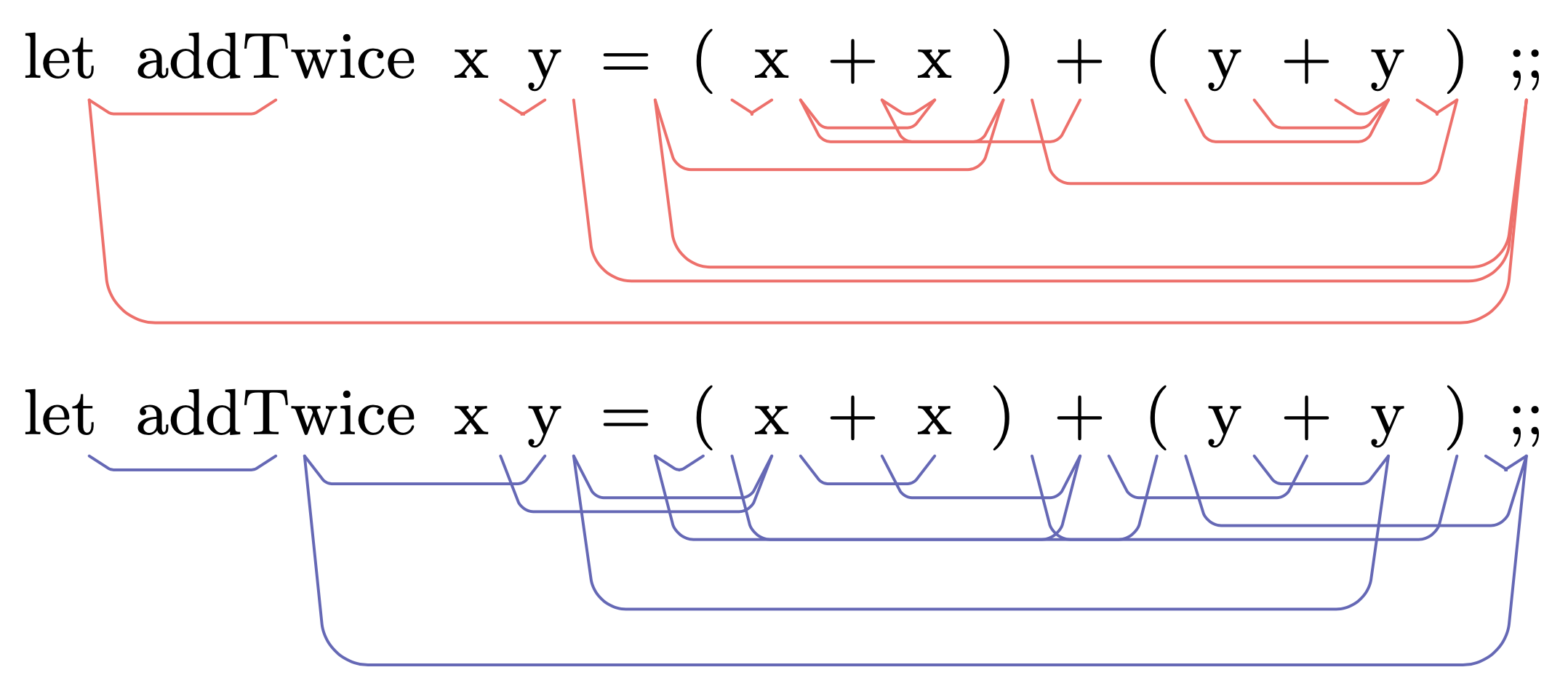}
         \caption{}
         \label{fig: ast_3}
    \end{subfigure}
    \vspace{-10pt}
    \caption{AST edge reconstruction from learned embeddings (edges in red are reconstructed by BERT Large model; edges in blue are reconstructed by Bidirectional LSTM model).}
    \label{fig: tree}
\end{figure}

\subsection{Explaining performance difference by the structural probe (RQ4)}

We use the structural probe to reconstruct structural information of programs based on both BERT's and Bi-LSTM's embedding representation. We hypothesize the difference in the ability of encoding structure information of programs could explain the performance gap between these two models. To gain some insights, we conduct a number of qualitative case studies. Fig~\ref{fig: tree} shows three examples of reconstructed AST of OCaml program using the structural probe.

Although the reconstructed ASTs are not adequate, we observe AST reconstructed from BERT embeddings turns to have many more meaningful structures than the AST reconstructed from Bi-LSTM embeddings. To be more specific, in Fig~\ref{fig: tree}(a), BERT's AST connects edge \texttt{(+, 5)}, \texttt{(2, *)} and \texttt{(*, a)} whereas Bi-LSTM's AST only has one meaningful edge \texttt{(*, a)}. Similarly, in Fig~\ref{fig: tree}(b), BERT's AST has an edge \texttt{(+, y)} while Bi-LSTM's AST fails to do so. In example (c), BERT's AST has an edge \texttt{(+, x)} and \texttt{(+, y)} which Bi-LSTM's AST omits.
Compared to Bi-LSTM, BERT is able to encode richer structural information, which may explain the huge 54\% performance gap between these two models. 

\section{Related Work}
\label{sec: related work}


Some recent works on type error diagnosis, such as SHErrLoc~\cite{sherrloc} and Mycroft~\cite{mycroft}, aim to analyze a set of typing constraints to find their minimum weight subsets~\cite{PLDI}. A minimum weight subset means omitting this subset will make the remaining constraints satisfiable and the subterms yielding the minimum weight subset inherit the blame. However, this approach has a few disadvantages. First, they are limited in terms of language choice. Different languages tend to employ different type systems and constraints. Thus, an approach designed for one type system can be hard to transfer to others. Secondly, the weights assigned are based on researchers' prior knowledge of the most likely errors instead of the most likely mistakes in practice~\cite{sherrloc}. Moreover, constraint-based approaches could blame a number of locations equally without taking the author's intent into account. 

\
In contrast, data-driven approaches such as \nate employ machine learning models to learn to localize type error from a large data set. While constraint analysis is not mandatory, \nate's machine learning models require considerable feature engineering. To be more specific, \nate employs over 282 hand-designed features annotated by human experts which are then fed into machine learning models to make the final prediction. However, \nate and other data-driven approaches still suffer from some disadvantages mentioned above. Although \nate doesn't perform constraint analysis, feature engineering also requires prior knowledge, making it difficult to transfer to other type systems. In addition, data-driven methods may implicitly consider the programmer's intent when making predictions, but there is no guarantee that such intent can be understood by models. In our study, we also show that the accuracy metric of \nate can be problematic in certain conditions.

There are also approaches that provide instructions to help novice programmers debug. Seidel et al. creates a dynamic model that generates counterexample witness inputs to show how the program goes wrong~\cite{dynamic}. When given a function with type errors, the algorithm symbolically executes the program and synthesizes witness the wrong values. Then the procedure is extended to a graph that shows the witness execution. Experimental results suggest their algorithm can generate witnesses 88\% of the time and in these successful programs, the algorithm yields counterexample successfully 81\% of the time. The advantage of this algorithm is that by using graphs and counterexamples, students can learn how to write code easily and understand the logic of the programming language. However, people who are familiar with the language but not that skilled, do not need such detailed suggestions. All they need is the precise location of the error. Chen et al. develops a type debugging system that asks programmers to provide type specifications during the debugging process and then generate suggestions that help to fix the type error~\cite{guided}. The advantage of this system is instead of aiming exclusively at the removal of type errors, it collects user feedback about result types to give useful suggestions, which include almost all possible corrections. This will help novices to debug more easily as they only need to choose from options given by the system. However, to achieve their goal, the authors systematically generate all potential type changes, which, when compared to our model, is more time-consuming in construction and needs more human judgment to make corrections.


There are also approaches which adopt SAT and SMT solvers to solve the type error localization. Pavlinovic et al. designs an algorithm that finds all minimum error sources, where 'minimum' is defined in terms of a compiler-specific ranking criterion~\cite{minimum}. With these error sources, a compiler is able to offer more useful reports. Then the authors try to reduce the search for minimum error sources to an optimization problem by implementing weighted maximum satisfiability modulo theories (MaxSMT). In this way, they leverage SMT solvers, making it easier to extend to multiple type systems and abstract from the concrete criterion that is used for ranking the error sources. The evaluation results on existing OCaml benchmarks for type error localization are also quite promising. In another work~\cite{PLDI}, Jose et al. aims to reduce the error localization problem to a maximal satisfiability problem (MAX-SAT), which finds the maximum number of clauses that are simultaneously satisfied by an assignment. Three steps are involved when an error should be reported. First, it encodes the denotation of a bounded unrolling of the program to a boolean formula. Then they construct an unsatisfiable formula for the failing program execution. In the last step, a MAX-SAT solver is used to find the largest set of clauses that can be satisfied at the same time, after which they output complement set as result, which is treated as potential locations of type error. Experimental results suggest the algorithm can find a few lines of code that are probable to be blamed for type error. Compared with our algorithm, the location it gives is too general. For novices, it is difficult for them to find the precise location of the type error when given such a large span of possible locations.

There are some other works that aim to diagnose the root causes of programs with typing errors. Chitil et al. uses a compositional type explanation graph created based on the Hindley-Milner type system~\cite{Chitil}. More specifically, this work relies on structural type information such as trees with principal typings. Tsushima et al. builds a type debugger without implementing any dedicated type inferencer~\cite{Tsushima}. The type debugger avoids re-implementing an independent type inference algorithm by leveraging the compiler’s type inference engine. In contrast to their work, we train natural language models to capture patterns in code changes. Our models do not require additional information beyond the code and can predict multiple error locations simultaneously. Our approach provides an orthogonal angle for (novice) type error diagnosis, and we believe that incorporating explicit type information can further improve our current approach.



\section{Conclusions and Future Work}
\label{sec: conclusion}
Many techniques have been developed to address the type error localization problem. While most of them employ static analysis on programs such as \sherrloc, \nate's success suggests that data-driven methods are also promising. Our experimental results suggest transformer-based language models outperform the state-of-art \nate and \sherrloc under a stricter yet more realistic accuracy metric.

Although being a black-box model, we show that language models can encode structural information of programs which may explain their better performance. Moreover, our models simply view a program as a sequence of tokens, thus they do not rely on any special knowledge of OCaml. It is the large amount of data (programs in our context) that plays an essential role in the performance. Since no feature engineering and constraints analysis are required, our approach can be transferred to other programming languages easily. We plan to investigate the effectiveness of our model on new languages like Go and Rust in the future. Through experiments, we identify several factors which help improve model accuracy such as size and positive transfer. We believe these factors may also be beneficial to solving other programming language-related tasks using language models.

In this work, our approach treats programs as natural language texts. This, however, fails to utilize the structural information of programs. Although we show language models could encode some structures, it is unclear how the encoded structural information leads to the final prediction. In contrast, constraint-based approaches such as \sherrloc take advantage of structural information and have much better interpretability. We plan to explore how to combine language models and the structural information of programs in our future work.

\section*{Acknowledgement}
We thank the anonymous reviewers for their insightful comments. This work was supported, in part, by
Individual Discovery Grants from the Natural Sciences and Engineering Research Council of Canada,
the Canada CIFAR AI Chair Program,
and Social Sciences and Humanities Research Council (SSHRC).



\clearpage
\bibliographystyle{splncs04}
\bibliography{refs}

\end{document}